\documentclass[preprint,12pt]{elsarticle}

\usepackage{graphicx}
\usepackage{amssymb}
\usepackage{amsmath}
\usepackage{amsfonts}
\usepackage{amssymb}
\usepackage{color}
\usepackage[cp1250]{inputenc}
\usepackage[english]{babel}
\usepackage{polski}
 \newtheorem{theorem}{Theorem}
\journal{}

\begin{document}

\begin{frontmatter}



\title{The intensity of the random variable intercept in the sector of negative probabilities.}

\author[cor5]{Marcin Makowski\corref{cor1}}
\ead{makowski.m@gmail.com}
\author[cor5]{Edward W. Piotrowski\corref{cor2}}
\ead{qmgames@gmail.com}
\author[js1]{Jan S\l{}adkowski\corref{cor3}}
\ead{jan.sladkowski@us.edu.pl}
\author[js1]{Jacek~Syska\corref{cor4}}
\ead{jacek.syska@us.edu.pl}
\address[cor5]{Institute of Mathematics, University of Bialystok, Pl 15245 Bia\l{}ystok, Cio\l{}kowskiego~1M, Poland}

\address[js1]{Institute of Physics, University od Silesia, Pl 40007 Katowice, Uniwersytecka 4, Poland}
\begin{abstract}
 We consider properties of the  measurement intensity $\rho$ of a random variable for which the probability density function represented by the corresponding Wigner function attains negative values on a part of the domain. We consider a simple economic interpretation of this problem. This model is used to present the applicability of the method to the analysis of the negative probability on markets where there are anomalies in the law of supply and demand (e.g.~Giffen's goods). It turns out that  the new conditions to optimize the intensity $\rho$ require a new strategy. We propose a strategy (so-called $\grave{a}$ rebours strategy) based on the fixed point method and explore its effectiveness.
\end{abstract}


\end{frontmatter}


\section{Introduction}
\label{intro}
We discuss the properties of the intensity $\rho$ of measurement of a real random variable $X$ 
with
density distribution
 $$pdf(p):=\frac{W(p,q=const.)}{\int_{-\infty}^{\infty} W(p,q=const.)dp}$$ restricted to domain $p\geq a$. $W(p,q)$ is the Wigner distribution of a certain quantum state for which function $W(p,q)$ is not positively defined,
where $q$ and $p$ are conjugate variables.
The definition of the functional $\rho$ is as follows:
\begin{equation}\label{first}
\rho \equiv \rho(pdf,a):= \frac{-\int_{-\infty}^{-a} p\cdot pdf(p)dp}{1+\int_{-\infty}^{-a}pdf(p)dp}\,.
\end{equation}
It is naturally related to the intensity of measurements in the context of transactions (buying/selling). Here parameter $a$ denotes the price below which a seller will always decide not to sell a given good, and the above quotient represents the rate of return in one cycle (buying/selling).  It is assumed that the duration of consecutive buying and selling cycles is a random variable and quotient (\ref{first}) expresses, in the language of probability density, the quotient of the average value of the profit rate in a whole cycle divided by the average duration time of this cycle. Details of the model and derrivation of formula (\ref{first}) are given in Section \ref{sec:1}.

The properties of (\ref{first}) were studied previously in classic and quantum market games \cite{fix,fix2,g} and in information theory context \cite{rQ, rT, sy}. We envisage that  the proposed  approach  has potential interesting applications in statistics (parameter estimation) \cite{est}.

The classical definition of a random variable implies monotonicity  of its distribution function. This fact leads to an interesting property of the corresponding intensity $\rho$: i.e. this function has a (global) maximum at its fixed point \cite{fix}. The assumption that the function $pdf(x)$
is non-negative is used in the proof of this property. It is therefore interesting to check how extreme properties
of
$\rho$ will change when this assumption is violated (the so-called negative probability).
For the purpose of the discussion an old interpretation of negative probabilities is adopted. Such problems have been known since the times of Captain Robert Giffen ('40s of of the 19th century \cite{GG,giff}) who was the first to notice the existence of a non-monotonic market demand for the so-called Giffen goods (modern description of the discovery is the result of formulating supply/demand curves for the logarithms of prices).

Analysis of such cases of $\rho$ is limited to the first excited state of the quantum harmonic oscillator. This state plays a particular role in an information-theoretic measurement which entails
the analysis of
distributions that minimise the Fisher information function (see subjective supply and demand curves in \cite{rQ,rT}).
The new properties of $\rho$ that are characteristic of quantum models can, for example, be used as a test for the existence of states with negative probabilities.
They also suggest a strategy for maximising the profit earned on transactions involving Giffen goods (see Setion \ref{sec5}).
\section{Simple market model. Profit intensity}
\label{sec:1}
The model presented in this section has been thoroughly studied, see \cite{fix,g,rQ,rT} for more details. It is the basis for our further considerations. Let us consider the simplest possible market event of exchanging two goods
which we would call  asset and  money and denote by $\Delta$ and $\$$, respectively. The proposed model comprises of two moves. First move consists
in a rational buying of the asset $\Delta$ (exchanging $\$$ for $\Delta$).  The second move
consists
in a random
(immediate) selling of the purchased amount of the asset $\Delta$ (exchanging $\Delta$ for $\$$). Let $V_{\Delta}$ and $V_{\$}$ denote some given amounts of the asset and the
money, respectively. If at some time $t$ the assets are exchanged in the proportion
$V_{\$}$:$V_{\Delta}$ then we call the logarithmic quotation for the asset $\Delta$ the number
\begin{displaymath}
p_t\equiv\ln(V_{\$})-\ln(V_{\Delta}).
\end{displaymath}
\\
If the trader buys some amount of the asset $\Delta$ at the quotation $p_{t_1}$ at the moment $t_1$ and sells it at the quotation $p_{t_2}$ at the later moment $t_2$ then his profit will be equal to
\begin{displaymath}
r_{t_1,t_2}=p_{t_2}-p_{t_1}\,.
\end{displaymath}
Let the expectation value of a
random variable $\xi$ in one cycle (buying-selling
or vice versa) be denoted by $E(\xi)$. If $E(r_{t,t+T})$ and $E(T)$,
where $T$ is the length of the cycle,
are finite then we define the profit intensity for one cycle
\begin{equation}\label{profit}
\rho_t\equiv\frac{E(r_t,_{t+T})}{E(T)}\,.
\end{equation}
We make the following assumptions:
\begin{itemize}
	\item The rational purchase -- a purchase bound by a fixed withdrawal price $-a$ that is such a logarithmic quotation for the asset $\Delta$,  above which the trader gives the buying up. A random selling can be identified with the situation when the withdrawal price is set to $-\infty$.
	\item Stationary process --
probability density  $pdf(p)$ of the random variable $p$ (the logarithmic quotation) does not depend on time.
\item $E(p)=0$\,. \footnote{It is sufficient to know the logarithmic quotations up to
arbitrary constant because what matters is the profit and profit is always a difference
of quotations.}
\end{itemize}
Let $x$ denote the probability that the rational purchase would not
occur:
\begin{displaymath}
x:=E([p>-a]),
\end{displaymath}
where
$$[P]=\left\{ \begin{array}{rl} 1 & \textrm{if P is true}\\0  & \textrm{otherwise}\end{array}\right .$$
The expectation value of the rational purchase time is equal to
\begin{displaymath}
	(1-x) +2x(1-x)+3x^2(1-x) +4x^3(1-x)+\cdots=\frac{1}{1-x}.\nonumber
\end{displaymath}
Therefore the mean length of the whole cycle is given by
\begin{displaymath}\label{czas}
	E(T)=1+E([p\leq -a])^{-1}.
\end{displaymath}
The logarithmic rate of return for the whole cycle is
\begin{displaymath}
r_{t,t+T}=-p_{\rightarrow\Delta}+p_{\Delta\rightarrow}
\end{displaymath}
where the random variable $p_{\rightarrow\Delta}$(quotation at the moment of purchase) has the
distribution restricted to the interval $(-\infty,-a]$:
\begin{displaymath}\label{ppp}
	pdf_{\rightarrow\Delta}(p)=\frac{[p\leq -a]}{E([p\leq -a])}\,pdf(p)\,.
\end{displaymath}
The random variable $p_{\Delta\rightarrow}$(quotation at the moment of selling) has the probability density $pdf(p)$.

After simple calculations (see \cite{fix} for details), using the formula (\ref{profit}), we obtain the expectation value of the profit after the whole cycle:
\begin{equation}
\rho(a)=\frac{-\int_{-\infty}^{-a} p\cdot pdf(p)dp}{1+\int_{-\infty}^{-a}pdf(p)dp}.
\end{equation}
It is easy to prove the following result \cite{fix}:
\begin{theorem}
The maximal value $a^{*}$ of the function $\rho(a)$ lies at a fixed point of this function ($\rho(a^{*})=a^{*}$). Such point exists, is uniquely determined and positive.
\end{theorem}
This theorem leads us to an interesting conclusion. The trader should
fix
the withdrawal price below the mean quotation so that the difference is the profit expected during a mean buying-selling cycle. Such a procedure is optimal.
\section{Demand and supply curves}
Supply and demand theory is an economic model of price in a market. Supply curve is a graphic representation of the relationship between product price and quantity of product that a seller is willing and able to supply. While demand curves is a graphic representation of the relationship between product price and the quantity of the product demanded. The study of these dependencies lies at the basis of economic. The two basic laws of supply and demand are:
\begin{itemize}
	\item As the price of a product increases, quantity demanded falls; likewise, as the price of a product decreases, quantity demanded increases.
	\item As the price of an item goes up, suppliers will attempt to maximize their profits by increasing the quantity offered for sale.
\end{itemize}
So,  demand and supply are  monotonic functions of prices. There is an interesting probabilistic interpretation of the supply and demand curves \cite{fix,rQ,rT}. Let us consider the functions:
\begin{eqnarray}
F_s(x)=\int_{-\infty}^{x}\eta_1(p)dp\,\,\,\,\,\,\,\,\,\,\,\,\,\,\,(supply)\,,\\
F_d(x)=\int_{x}^{\infty}\eta_2(p)dp \,\,\,\,\,\,\,\,\,\,\,\,\,\,\,(demand)\,,
\end{eqnarray}
where $\eta_1$, $\eta_2$ are appropriate probability density function, in general case different due to various properties of the market (monopoly, specific
market regulations, taxes).
The value of the supply function $F_s(x)$ is given by the probability of the purchase of a unit at
the price $\leq e^x$ (and analogously in the case of demand). More details about this interpretation of supply and demand curves can be found in \cite{rQ,rT}.
\section{Exceptions to the law of demand/supply}
As we mentioned above, the demand and supply is monotonic function of prices. In the probabilistic interpretation it means that probability
density is non-negative. This assumption  is used in the proof of Theorem 1.
It turns out that violations of the law of demand/supply are sometimes observed on markets. One of the most well-known example is the already mentioned Giffen's goods \cite{GG,giff}. It is a product that people consume more  as its price rises. In our convention, this situation corresponds to the resignation from the assumption that probability density is non-negative. Let us see how it affects the profit intensites and strategies of traders for this type of market. For this, we introduce Wigner quasiprobability distribution \cite{rW}:
\begin{displaymath}
W(p,q)=\frac{1}{\pi} e^{-p^2 - q^2} (-1 + 2 p^2 + 2 q^2).
\end{displaymath}
It is the function describing the first excited state for one-dimensional harmonic oscillator. In our economic interpretation, it corresponds to the most basic (single) variant of violation of the supply/demand laws. Thus, the probability density is of the form
\begin{equation}\label{densty}
pdf(p,q=const.):= \frac{W(p,q=const.)}{\int_{-\infty}^{\infty}W(p,q=const.) dp}\,.
\end{equation}
 The condition $ q = const. $ can be interpreted as a specific market property, which indicates the scale of violation of the supply/demand laws. The impact of this constant on the effectiveness of the strategies will be to discuss
  in the rest of the work.

\section{Example}

\begin{figure}[!h]
\begin{center}
\includegraphics[width=9cm]{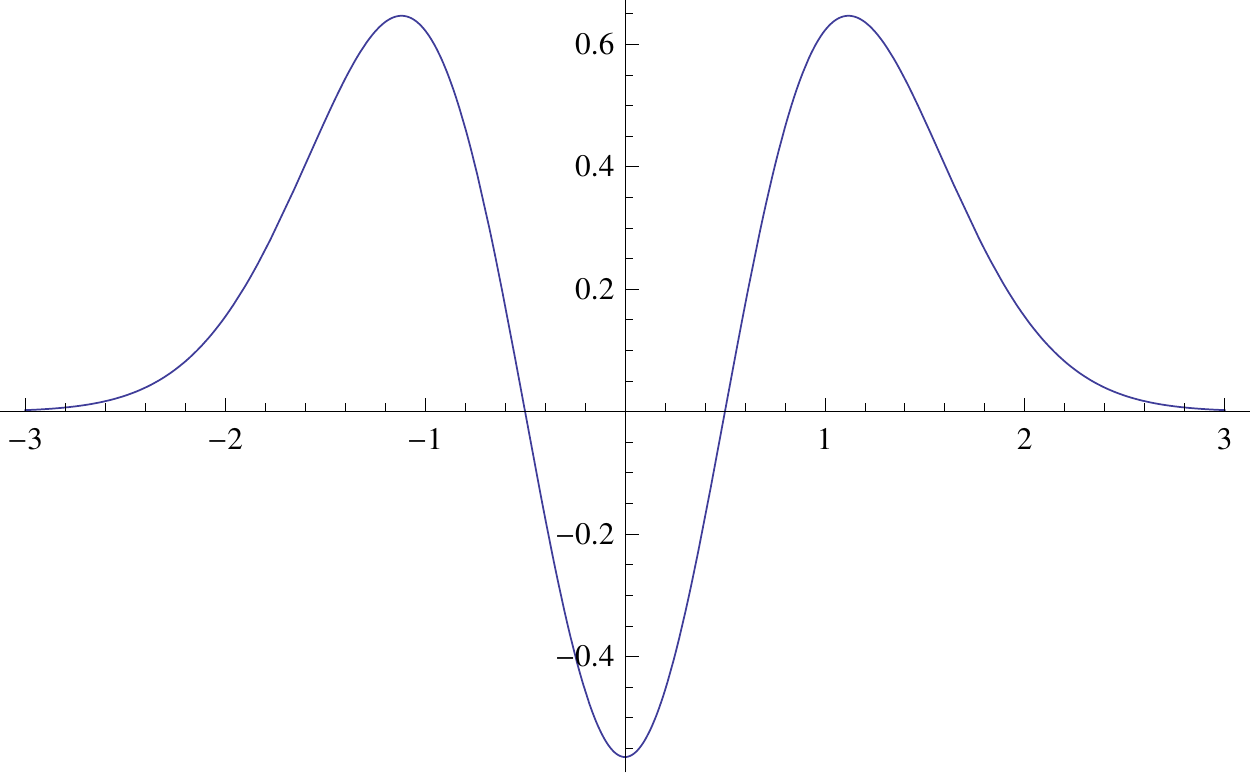}\caption{Graph of the pdf(p,q=0.5) function.}\label{pdf}
\end{center}
\end{figure}
Let us consider an example of how the situation of a trader changes in the presence  of violations of the law of demand/supply. Let us assume that $q=0.5$. The corresponding
$pdf$
is plotted in Fig.~\ref{pdf}.

\begin{figure}[!h]
\begin{center}
\includegraphics[width=9cm]{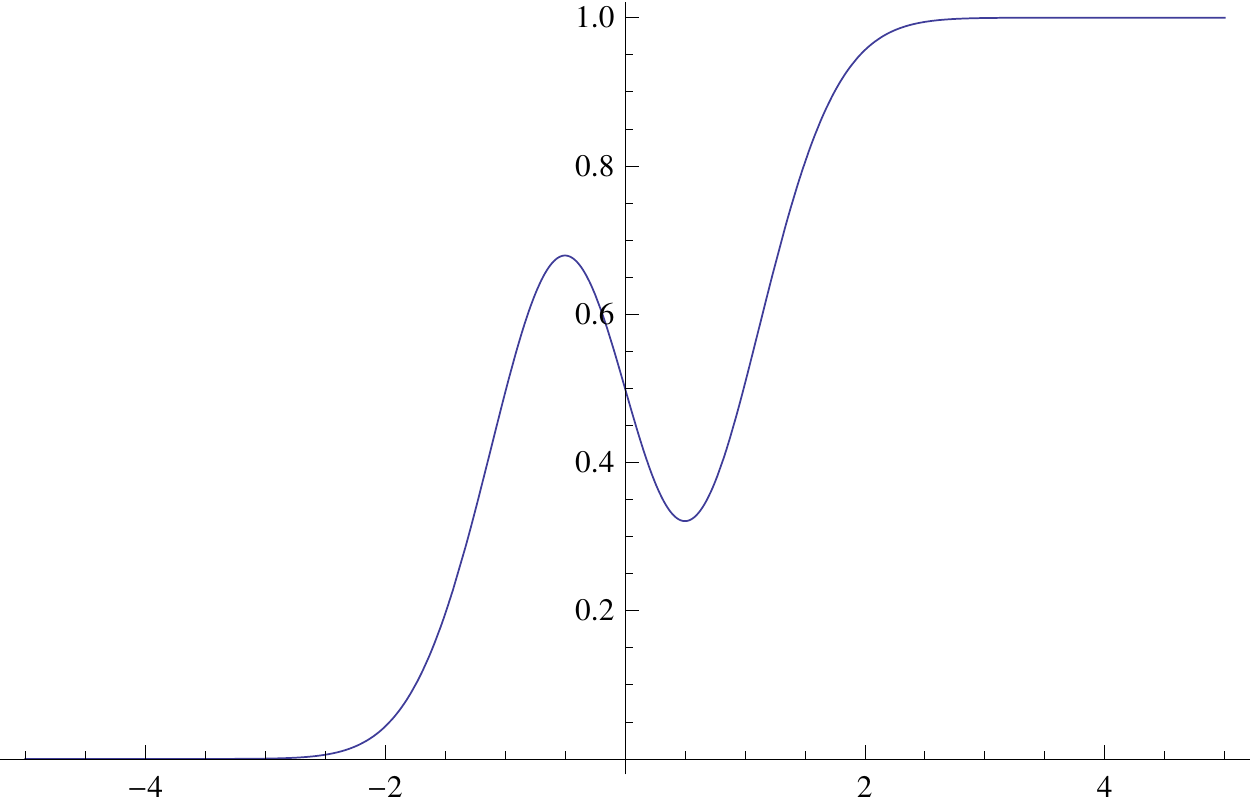}\caption{Graph of the cdf(p,q=0.5) function.}\label{cdf}
\end{center}
\end{figure}

This function is negative on some interval (if $q\in(-\frac{1}{\sqrt{2}},\frac{1}{\sqrt{2}})$). This results in that the cumulative distribution function (supply function)  is not monotonic (analogously for the  demand function). It can be observed in Fig.\ref{cdf}.

\begin{figure}[!h]
\begin{center}
\includegraphics[width=9cm]{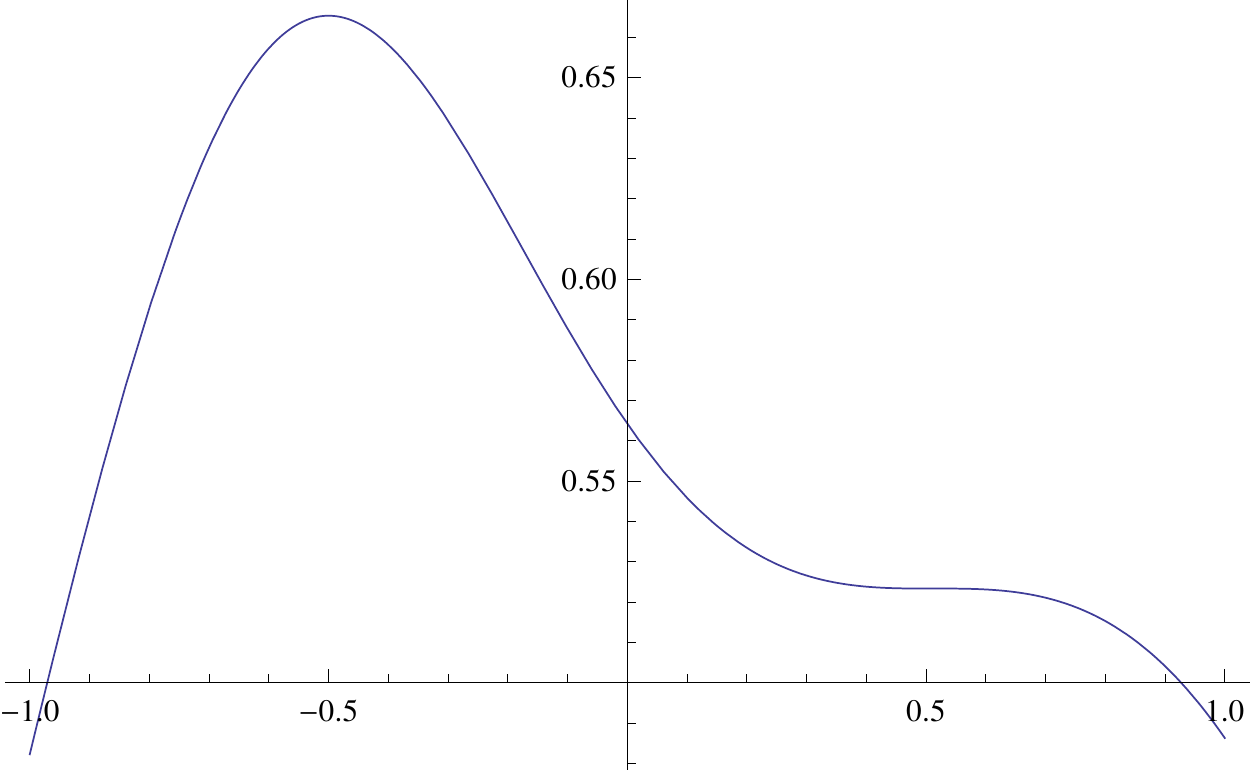}\caption{An example of $\rho(a)$ function for $pdf(p,q=0.5)$.}\label{rho}
\end{center}
\end{figure}

Under the foregoing assumptions we obtain profit intensity function $\rho(a)$, see Fig.\ref{rho}.

As we can see fixed point theorem does not work here! We have maximum in fixed point but it is not
 the global one.
\section{$\grave{A}$ rebours strategy}\label{sec5}
We are referring here to a realistic economic situation, therefore, we assume  nonnegativity of $F_s$ and $F_d$. In this case, the global maximum of the expectation value of the profit 
$\rho(a)$ 
is for
\begin{equation}\label{wmax}
a=-\sqrt{\frac{1-2q^2}{2}}\,,
\end{equation}
and the  minimum for
\begin{equation}\label{wmin}
a=\sqrt{\frac{1-2q^2}{2}}\,,
\end{equation}
where $|q|\in D=[\alpha,\frac{1}{\sqrt{2}}-\epsilon]$.\footnote{If  $q\in\mathbf{R}\setminus D$ we get the global maximum at the fixed point of $\rho(a)$ function.}

The value of the parameter $ \alpha \approx 0.38375 $ corresponds to the limit of the parameter $ q $ above which
functions $ F_s $ and $ F_d $ achieve non-negative values. 

The section $ (\frac{1}{\sqrt{2}} - \epsilon, \frac{1}{\sqrt{2}}) $, where $ \frac{1}{\sqrt{2}} - \epsilon \approx 0.69590 $, can be called \textsl{the non-classical sector of effectiveness for classical optimization}. This is an area of these values $ q $ for which we observe derogation from the supply/demand laws (where the functions $F_s$ and $F_d$ are not monotonic) but here the thesis of the Theorem 1 is true, i.e.~the function $ \rho(a) $ has the global maximum at its fixed point.
This area is only approx.~$3.47 \%$ of the interval $ [\alpha, \frac{1}{\sqrt{2}}] $. In addition, there are values of $ q $ close to  $ \frac{1}{\sqrt{2}} $, after which the functions $F_s$ and $F_d$ becomes monotonic, the market behaves according to the supply/demand laws. Hence for  $q$ from this sector we observe violations of the supply/demand laws. They are so small that they practically do not force the player change strategy.

Note that by adequately adjusting the parameter $q$ supply (or demand) curve, we can cause that the player who adopts a classical strategy (fixed-point search) will demonstrate a small intensity of profits. This is well illustrated by the example given above, see Fig.~\ref{rho}. It is characteristic that the value of the function $ \rho $ attains its local minimum 
 which 
is close to a local maximum in the fixed point. Moreover, the value (\ref{wmax}) at which we have global maximum of $\rho$ has  opposite sign to the value (\ref{wmin}), for which we obtain  minimum.

\begin{figure}[!h]
\includegraphics[width=6.5cm]{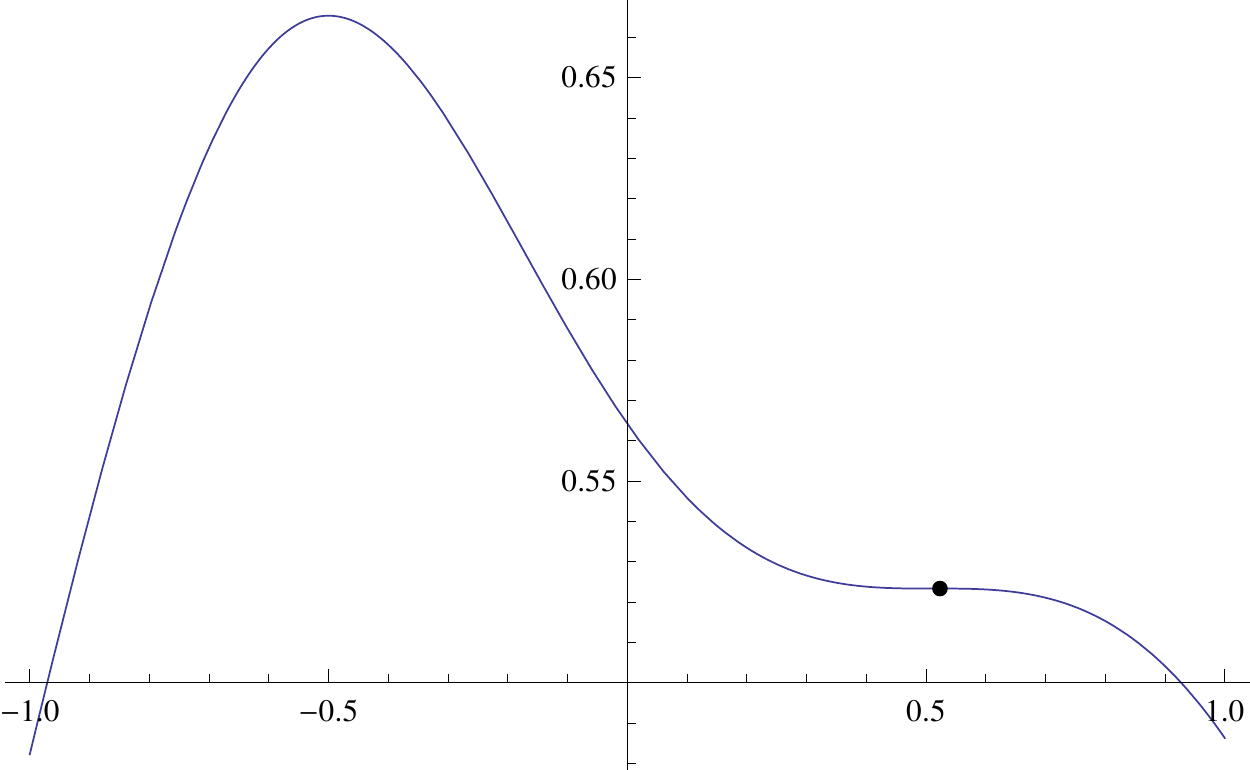}\,\,\,\,\,\,\,\,\,\,\,\,\,\,
\includegraphics[width=6.5cm]{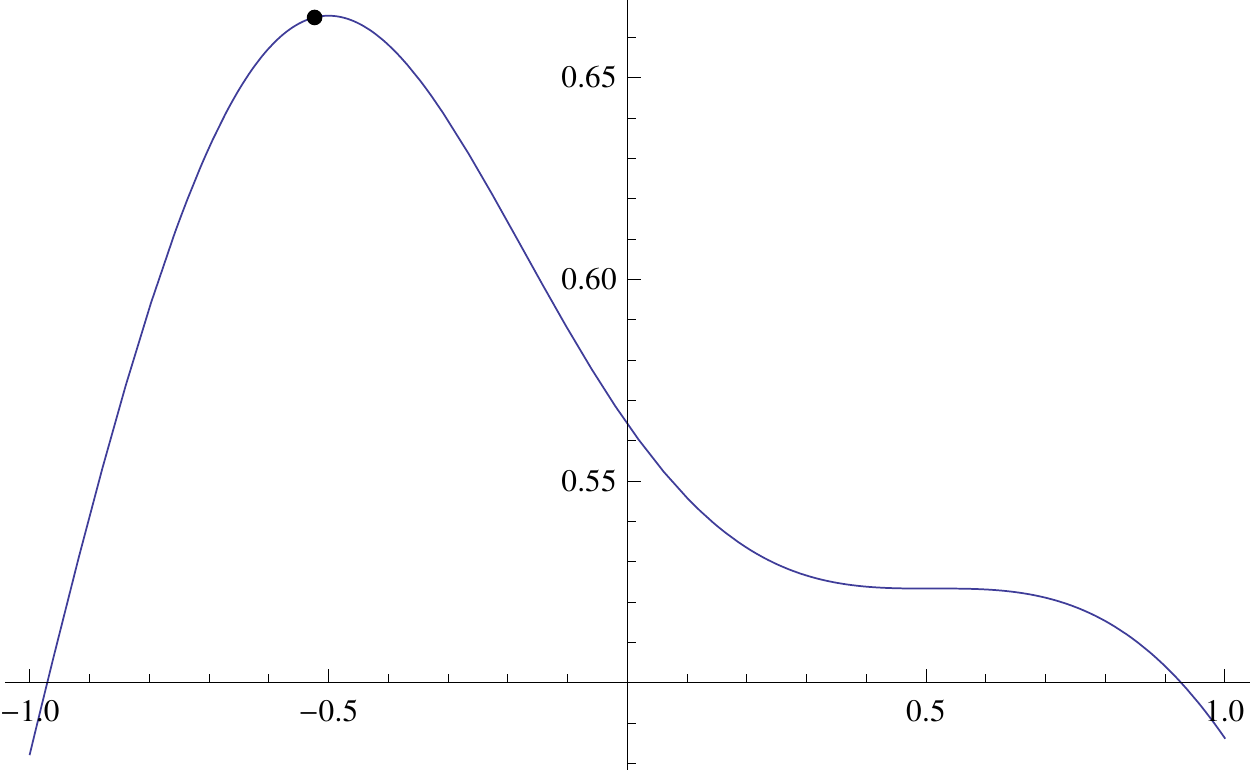}\caption{The value of the profit intensity. Using classical fixed point strategy (left point) and $\grave{a}$ rebours strategy (right point).}\label{reb}
\end{figure}

This suggests a new type of strategy:
\begin{quote}\emph{locate the fixed point $a^*=\rho(a^*)$ and then take $-a^*$ as a new withdrawal price}\/.\end{quote}
Let's call it \textsl{$\grave{a}$ rebours strategy}. Doing so allows the player to reach the intensity of her/his profits close to its maximum. A plot of  results of using this strategy with $q=0.5$ is presented in Fig.~\ref{reb}.
The effectiveness\footnote{Understood as getting the profit intensity $\rho$ close to its maximal value.} of the $\grave{a}$ rebours strategy varies depending on the value of $q$ (the highest is in the vicinity of $q=0.5$). Table~\ref{tab} 
shows 
(for some $q$) the values of profit intensity for discussed strategies and their maximum values.
 \begin{table}[h!]\caption{The value of $\rho$ function: global maximum (A), fixed point strategy (B), $\grave{a}$~rebours strategy (C).}\label{tab}
\begin{tabular}{|l|l|l|l|l|l|l|l|}
\hline
&\multicolumn{7}{|c|}{The value of the \textsl{q} parameter } \\\hline
q= & 0.38375  & 0.50000 & 0.6000 & 0.65000 & 0.68000  & 0.69590 & 0.70700\\\hline\hline
 A  & 1.34619 & 0.66531 & 0.47086 & 0.41719 & 0.39325 & 0.38267 & 0.37618\\\hline
 B  & 0.67323 & 0.52335 & 0.43861 & 0.40697 & 0.39063 & 0.38267 & 0.37738\\\hline
 C  & 1.32678 & 0.66488 & 0.46954 & 0.41395 & 0.38799 & 0.37596 & 0.36817\\\hline
\end{tabular}
\end{table}
It can be seen that with the increase in $q$ 
the usefulness of the classic strategy 
increases
(it gives even better results for  $q$ close to the limit values,
 for example $q = 0.68$ in Table \ref{tab}). However, when comparing both of these strategies on all admissible  values of $ q $, $ \grave{a} $ rebours strategy gives better results in most of its parts. Moreover, even when the classical strategy is effective, if it is used in the same conditions $\grave{a}$ rebours strategy produces results slightly deviating from the maximum. The same is not true for  the classical strategy (see $q=0.38375$ in Table \ref{tab}).

\section{Conclusion}
The search for optimal solutions and  fixed points of mappings are among
the basic tasks of mathematics (e.g. Brouwer theorem, Banach contraction mapping principle, etc.). Such problems were from the beginning closely related to their applications in fields such as game theory and economics \cite{Fg}, statistics \cite{est} and many other \cite{fa,Sb}. In this paper we have examined the measurement intensity of a random variable in 
the
sector of negative probabilities. It turns out that in such circumstances the values close to the maximum can be obtained for the argument of intensity $\rho $ which has the opposite sign to the fixed point $\rho$. An interesting property of the discussed model is the presence a non-classical sector of effectiveness for classical optimization. This means that a fixed point optimization model is insensitive to small fluctuations of probability density distributions in the area of non-classical
probabilities (i.e. negative probabilities). This results in stability of classical strategies. This unique result is worth of exploration in a wider context.

Transactional interpretation of the model presented in the article  shows that the analysis of negative probabilities can also be used in studies of  markets with violations of the classical laws of supply and demand. We envisage 
that 
this may lead to  interesting research  within intensively developed in recent years field of econophysics.

\section*{Acknowledgments}
This work was supported by the Polish National Science Centre under the project
number \textbf{DEC-2011/01/B/ST6/07197}.\newpage

\end{document}